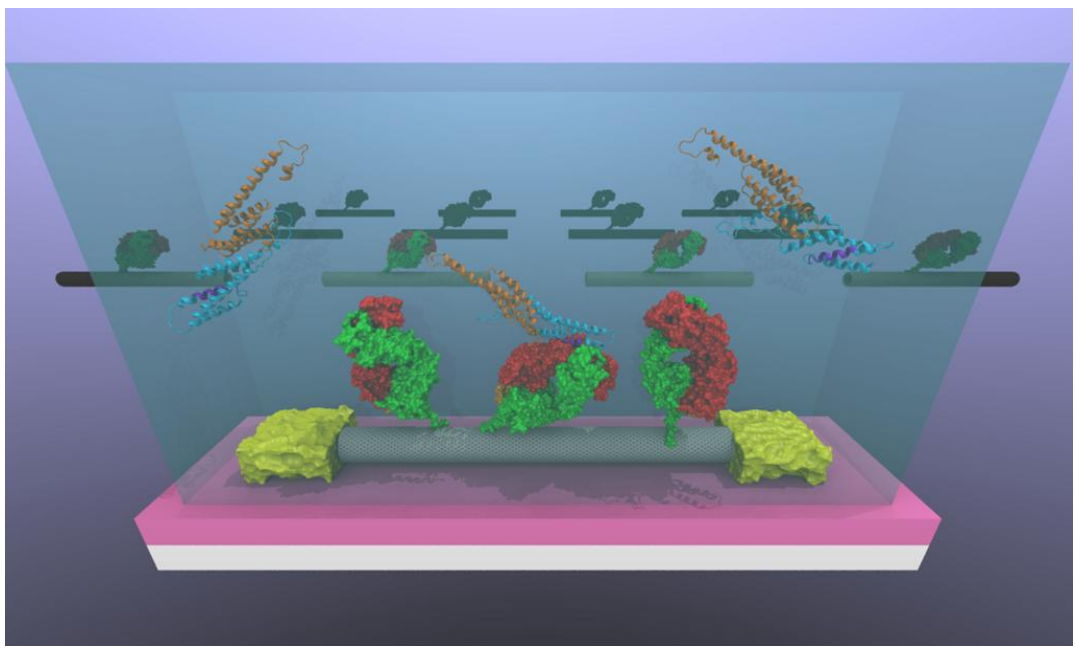

# Hybrids of a Genetically Engineered Antibody and a Carbon Nanotube Transistor for Detection of Prostate Cancer Biomarkers

Mitchell B. Lerner[1], Jimson D'Souza[2], Tatiana Pazina[2], Jennifer Dailey[1], Brett R. Goldsmith[1], Matthew K. Robinson[2], A.T. Charlie Johnson[1]

[1] Department of Physics and Astronomy, University of Pennsylvania, 209 S. 33rd St., Philadelphia, PA
     19104
[2] Developmental Therapeutics Program, Fox Chase Cancer Center, 333 Cottman Avenue, Philadelphia,
     PA 19111

## Abstract-

We developed a novel detection method for osteopontin (OPN), a new biomarker for prostate cancer, by attaching a genetically engineered single chain variable fragment (scFv) protein with high binding affinity for OPN to a carbon nanotube field-effect transistor (NTFET). Chemical functionalization using diazonium salts is used to covalently attach scFv to NT-FETs, as confirmed by atomic force microscopy, while preserving the activity of the biological binding site for OPN. Electron transport measurements indicate that functionalized NT-FET may be used to detect the binding of OPN to the complementary scFv protein. A concentration-dependent increase in the source-drain current is observed in the regime of clinical significance, with a detection limit of approximately 30 fM. The scFv-NT hybrid devices exhibit selectivity for OPN over other control proteins. These devices respond to the presence of OPN in a background of concentrated bovine serum albumin, without loss of signal. Based on these observations, the detection mechanism is attributed to changes in scattering at scFv protein-occupied defect sites on the carbon nanotube sidewall. The functionalization procedure described here is expected to be generalizable to any antibody containing an accessible amine group, and to result in biosensors appropriate for detection of corresponding complementary proteins at fM concentrations.

## Introduction

Prostate cancer (CaP) is a major public health issue as the most commonly diagnosed cancer and second leading cause of cancer deaths among American men[1]. Detection of early-stage cancer often results in successful treatment, with long-term disease-free survival in 60-90% of patients. One methodology for early disease detection is biomarker sensing and quantification through use of specialized assays. Biomarkers of cancer are molecular or tissue-based signatures of disease that provide either diagnostic or prognostic insight into disease etiology and/or progression. In the case of CaP, levels of Prostate Specific Antigen (PSA) in patient serum above 4 ng/mL[2] have traditionally been used as a predictive biomarker of disease. However, PSA tests are prone to both false positives and false negatives, causing healthy men to undergo unnecessary medical procedures as well as men with cancer to go undiagnosed[3]. This has resulted in the utility of PSA as a broad-based screening tool being called into question and supports the need for identification of additional biomarkers of CaP. One potential new biomarker of CaP is Osteopontin (OPN)[4]. OPN is a proinflammatory cytokine that regulates bone homeostasis through its effects on osteoclast function[5]. OPN is being investigated both as a therapeutic target and as a biomarker for diseases such as arthritis and cancer[6,7]. Progression of CaP, as well as other cancers, is often associated with metastasis to bone. OPN is thought to play a role in the metastatic process and development of metastatic disease correlates with increased serum levels of OPN[8]. Monoclonal antibodies (mAbs) specific for OPN, such as the 23C3 mAb , represent promising therapeutic and diagnostic agents[9,10]. Traditional protein detection methods such as ELISA are quite sensitive, but have proven problematic for quantification of OPN[11]. They also require pure samples, lengthy processing times, expertise in molecular biology, and could be expensive[12]. As a result, cost efficient, easy to implement immunosensors with comparable or better sensitivity than an ELISA assay would be highly desirable. Recently, it was reported that biosensors based on semiconducting nanowire field effect transistors (FETs) achieved detection limits for PSA of 5 pM for time domain measurement and 150 fM for frequency domain measurements that require more complex electronic instrumentation[13]. This motivated us to explore whether further sensitivity enhancements could be achieved through the use of carbon nanotube FETs tightly coupled to engineered antibody elements.

Carbon nanotube field effect transistors (NT FETs) provide a unique transduction platform for chemical and biomolecular detection[14-16]. Tailored and specific detection may be accomplished by chemically functionalizing the NT FET with a selected antibody to create a hybrid nanostructure. Proteins that bind to the hybrid alter the electrical properties of the NT FET via several mechanisms[17], allowing direct detection as a change in the transistor conduction properties such as threshold voltage or ON state current. Ease of fabrication, well-understood carbon surface chemistry, and fast electronic readout make NT FET-based hybrids desirable as sensors in either an antibody-antigen detection scheme or as a vapor sensor suitable for more complex system architectures characteristic of mammalian olfaction[18-21]. Large arrays of such devices could potentially assist medical diagnosis through simultaneous measurement of hundreds of biomarkers using a single small-volume sample[22-24].

The binding specificity inherent in mAbs makes them promising agents for use as both targeted therapies and as reagents for detecting biomarkers of disease. The murine mAb 23C3 recognizes a conserved, linear peptide ($^{43}$WLNPDP$^{48}$) within OPN. When administered to animals the 23C3 immunoglobulin (IgG) induced a therapeutic response in a collagen-induced model of arthritis[6] and led to its humanization (Hu23C3) through a CDR-grafting approach in preparation for transition into clinical trial[25]. Effective use of targeted therapies for the treatment of cancer and other diseases requires paired diagnostic tests to detect the presence or absence of relevant biomarkers.

In this study we have engineered the 23C3 mAb into a single-chain variable fragment (scFv) antibody[26]. ScFv antibodies are comprised of the variable heavy and variable light domains of the parental IgG, fused by a short peptide linker, and retain the antigen binding properties of the intact IgG. The 23C3 mAb was converted to an scFv in an effort to optimize its structure for use on NT FET-based biosensors for the detection of OPN in bodily fluids. Here we found that the 23C3 scFv retains its ability to bind OPN, which should allow it to be an effective diagnostic surrogate for the Hu23C3 therapeutic antibody. In addition, its small size compared to the parental IgG (25kDa and 150 kDa, respectively) is hypothesized to provide advantages over the parental IgG when used for production of NT FET-based biosensors. First, scFv can be expressed in a bacterial cell system that improves both the speed and cost of goods associated with their production as compared to traditional IgGs. Second, the scFv is comprised entirely of the antigen-binding domain of the 23C3. Thus, antigen binding by an immobilized scFv is hypothesized to bring OPN into closer proximity to the NTFET surface, as compared to what would occur with an intact IgG. This should result a greater impact on the electrical properties of the device and potentially improve the sensitivity of detection.

Here we report successful fabrication of NT FETs covalently functionalized with the 23C3 scFv, as evidenced by Atomic Force Microscopy and electronic measurements. We observed an antigen-specific, concentration-dependent sensor response to OPN in buffer with a detection limit below 30 fM. The response as a function of concentration was well fit by a model based on the Hill-Langmuir equation of equilibrium thermodynamics.

## Results and Discussion

NT FET devices were fabricated as described in the Methods section. Briefly, NTs were grown by catalytic chemical vapor deposition, and electrical contacts were patterned using a photolithographic technique that is optimized to leave a chemically clean NT sidewall[27]. Pristine $sp^2$-bonded carbon is chemically inert, so steps must be taken to create a defect on which to bind proteins (Figure 1). Diazonium salt treatment[28] was used to create $sp^3$-bonded sites along the nanotube terminated in a carboxylic acid group that was used for further attachment chemistry. Raman spectroscopy measurements showed an increase in the ratio of the intensity of the D band to the G band after diazonium salt treatment, consistent with the creation of covalent bonds to the NT sidewall (see Supplemental Figure 1 in the Supporting Information)[29].

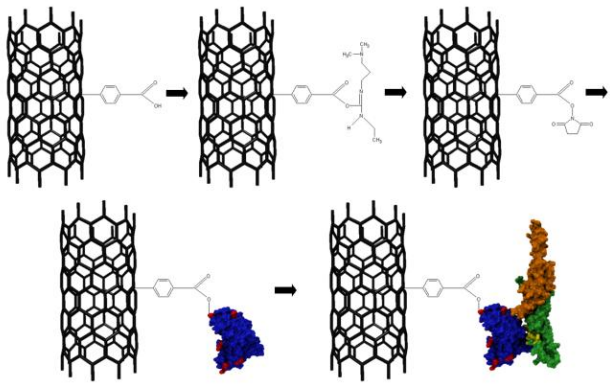

Figure 1: **Functionalization scheme for OPN attachment.** First, sp3 hybridized sites are created on the nanotube sidewall by incubation in a diazonium salt solution. The carboxylic acid group is then activated by EDC and stabilized with NHS. ScFv antibody displaces the NHS and forms an amide bond (surface amine-rich lysine residues responsible for this bond are depicted in red), and OPN binds preferentially to the scFv in the detection step. The OPN epitope is shown in yellow, and the C-and N-terminuses are in orange and green respectively.

The diazonium treatment step was followed by activation and stabilization of the attachment site with 1-ethyl-3-[3-dimethylaminopropyl]carbodiimide hydrochloride/sulfo- *N*-hydroxysuccinimide (EDC/NHS)[30]. Amine groups associated with lysine residues on scFv proteins were expected to displace NHS in the subsequent attachment step to form a covalent bond between the scFv and the NT through the phenolic linker. A careful washing procedure was employed to minimize scFv binding to the substrate. We expect that the chemical functionalization procedure used here for the scFv derived from the 23C3 mAb[31] may be generalized to broad classes of protein that have a free amine group on their exterior.

Preferential attachment of scFv antibodies to NTs was confirmed by Atomic Force Microscopy (AFM) (Fig. 2a). The linear density of attached scFv antibodies was found to be approximately 4-5 per μm of NT length. Line scans from the AFM images were used to calculate the size of ~180 features presumed to be scFv antibodies covalently bound to NTs and these data were used to form a histogram shown in Fig. 2b. The histogram had its maximum at ~ 2.5 nm, with secondary maxima at 5 nm and 7.5 nm. We attribute the primary maximum to attachment of single anti-OPN scFv molecules, and note that a molecular diameter of 2.5 nm is consistent with their mass of 25 kDa; the other maxima are attributed to small scFv aggregates.

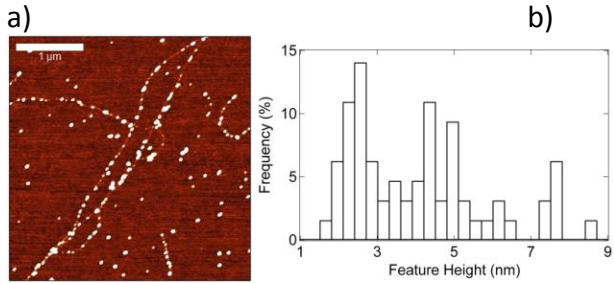

Figure 2: a) AFM image of scFv antibodies covalently bound to carbon nanotubes, with a typical density of 4-5 attachment sites per micrometer of nanotube length. Scale bar is 1 μm. Height scale is 7 nm. b) Histogram of the heights of the white features bound to the nanotube in Fig 2a shows a maximum at ~ 2.5 nm, consistent with the expected height of scFv antibodies. Secondary maxima at 5 nm and 7.5 nm are attributed to small aggregates of scFv antibodies. The total number of features analyzed is 180.

We measured the current-gate voltage (I-Vg) characteristic of an individual NT FET device after each chemical modification to monitor the effect of the chemical functionalization and confirm attachment of scFv antibodies (Fig. 3). The most sensitive parameters were the threshold voltage where the FET current decreases most sharply and the ON state current carried by the device for very negative values of the gate voltage. The diazonium treatment was found to shift the turn-off voltage shifted by about -3V, and it reduced the ON state current by 50-90%, consistent with increased electron scattering due to generation of defects on the NT sidewall. EDC/NHS treatment caused almost no change in the turn-off voltage and a slight decrease in the ON state current. ScFv attachment did not change the turn-off voltage and led to a small increase in the ON state current, attributed to a reduction in carrier scattering due to the presence of the protein compared to that associated with NHS/EDC activation.

To test the sensitivity of the device to exposure to OPN, a 10 μL droplet of PBS buffer containing OPN at a known concentration was placed on the sensor and left to incubate for 20 minutes in a humid environment, followed by careful washing with DI water. The 20-min incubation time we used should allow OPN molecules to diffuse distances of order 300 μm, which we expect is sufficient to establish equilibrium between bound and unbound species even at the lowest concentration used (1 pg/mL, equal to approximately 30 fM)[32, 33]. Each device tested was used for a single measurement at a fixed OPN concentration to avoid contamination of the samples. The ON state current proved to be very sensitive to the presence of OPN and showed a reproducible increase after exposure to OPN (Fig. 3). The response is reported as ΔI/I, the percentage increase in ON state current from the scFv functionalized state to the state following OPN exposure, which is found to account well for device-to-device resistance variations.

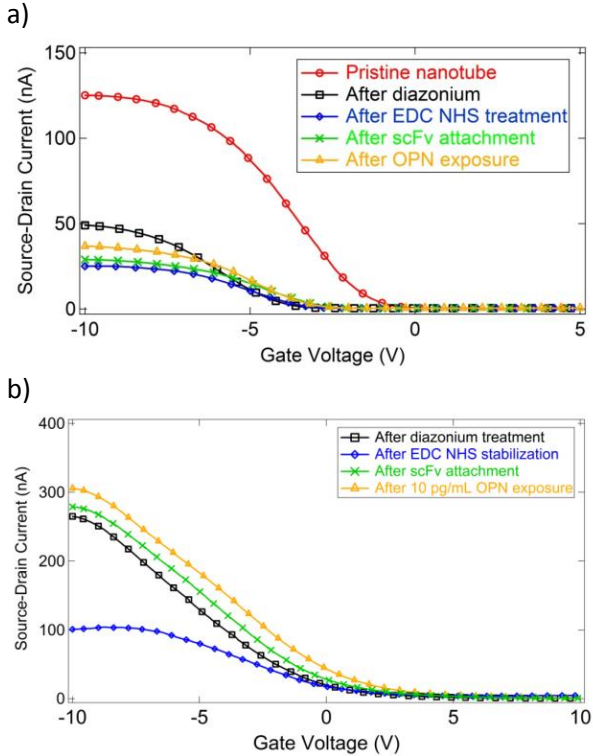

Figure 3: a) I-V$_g$ characteristics for a single device after successive functionalization steps. Exposure to OPN at a concentration of 30 ng/mL (912 pM) caused the ON state current to increase from 29 nA (green curve) to 37 nA (orange curve), an increase of 27%. b) Similar data for exposure to OPN at a concentration of 10 pg/mL (304 fM). The ON state current increases from 276 nA to 307 nA, an increase of 11%.

The device response varied systematically with OPN concentration as shown in Figure 4. Each data point in the plot was averaged data from 5-10 functionalized NT FET devices tested against a solution with the same concentration of OPN. To generate a fit to the data, we used a model where the device response was comprised of two additive components: an offset response, Z, due to incubation in pure buffer with no OPN and a response that is directly proportional to the probability that an OPN binding site is occupied. This motivated a fit based on a modified Hill-Langmuir equation describing ligand-receptor binding in equilibrium[34].

$$\frac{\Delta I}{I} = A \frac{(c/K_d)^n}{1+(c/K_d)^n} + Z$$

Here $c$ is the OPN concentration, $A$ is the response when all binding sites are occupied, Z is an overall offset to account for buffer response, $K_d$ is the dissociation constant, and $n$ is the Hill coefficient describing cooperativity of binding. The best fit to the data yielded an offset parameter $Z$ = 4%, dissociation constant $K_d$ = 564 pg/mL, maximum response $A$ = 27%, and $n$ = .201. The maximum response A and the offset parameter Z were constrained to values that were sensible based on the response data and buffer response respectively while n and $K_d$ were free to vary. The value of $K_d$ (1.7 x 10$^{-11}$ M) obtained by the NT FETs is somewhat lower than $K_d$ derived from SPR measurements (Supplemental Figure 1). This difference may be explained by the fact that SPR is performed under conditions of buffer flow that preclude re-binding of the OPN to the anti-OPN scFv during the dissociation phase measurements. In contrast, the static conditions used with the NT FETs may allow for re-binding, slowing the apparent off-rate and resulting in the lower apparent $K_d$. Interestingly, the values derived on the NT FET are equivalent to those predicted for the Hu23C3 mAb derived from ELISA-based methods[25]. The observed value n = .201 indicates negative cooperativity in the binding of OPN to the 23C3 scFv in the context of the NT FET biosensor. This could be due to inhibition of multiple OPN molecules binding to small aggregates or clusters of scFv due to steric hinderance. Taking into account the experimental noise, the inferred detection limit of the scFv-functionalized NT FET devices for OPN was 1 pg/mL, or 30 fM.

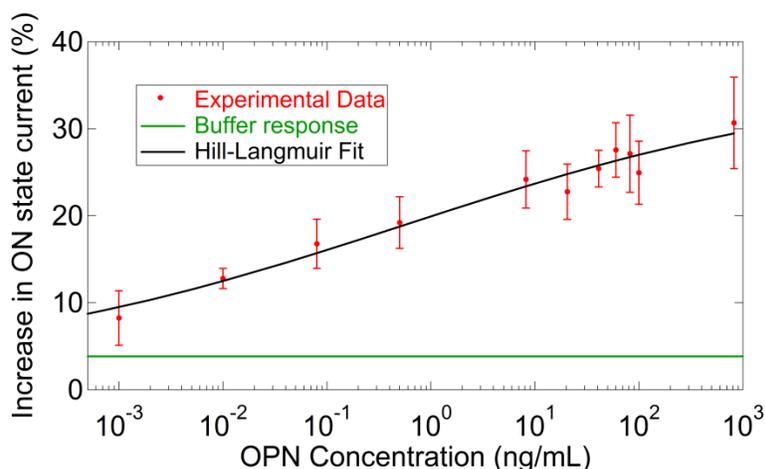

Figure 4: Measured sensor responses over a wide range of concentrations of osteopontin (OPN). The solid line is a fit using a modified Hill-Langmuir expression that includes an offset response of 4% due to the buffer itself (see main text). A clear signal is still present at OPN concentrations of 1 pg/mL.

The precise mechanism for the observed sensing response remains to be determined. Qualitatively, an increase in ON state current implies a reduction in carrier scattering upon OPN binding. One possible explanation is that charged sites on the scFv surface are neutralized by opposite charges associated with bound OPN, leading to reduced fluctuations in the electrostatic potential at the NT surface. Even in the absence of a quantitative understanding of the device response, our results provide strong evidence that the methods used here enabled attachment of engineered proteins to a NT FET while maintaining both the high quality electronic characteristics of the NT device and the chemical recognition functionality characteristic of the protein.

Control experiments were conducted where the scFv-functionalized sensor was incubated in a solution of bovine serum albumin (BSA) at high concentration (450 ng/mL) to approximate the effect of non-specific proteins that will be present in patient samples. The devices were found to give a null response of -0.24 % ± 3.87% (Figure 5), supporting the notion that the scFv antibodies bound to the NT sidewall retained their specificity for OPN. In a second test of antigen specificity, we functionalized NT FET devices with a different scFv based on the clinically validated therapeutic monoclonal antibody trastuzumab[35]. Trastuzumab, and the derived 4D5 scFv, binds to the HER2 receptor tyrosine kinase, which is implicated in breast cancer formation and progression[35]. Surface Plasmon Resonance measurements show that the 4D5 scFv binds to HER2 with a $K_D$ of 1.9 nM[36], equivalent to the binding affinity of 23C3 for OPN. As predicted the 4D5 scFv exhibits no binding to OPN when analyzed by SPR (Supplemental Figure 3). As detailed in Figure 5, 4D5-functionalized devices show no response after exposure to OPN at 100 ng/mL, consistent with the SPR data. These control experiments suggest that anti-OPN 23C3 scFv-functionalized carbon nanotube sensors exhibit a high level of specificity for OPN, the target ligand.

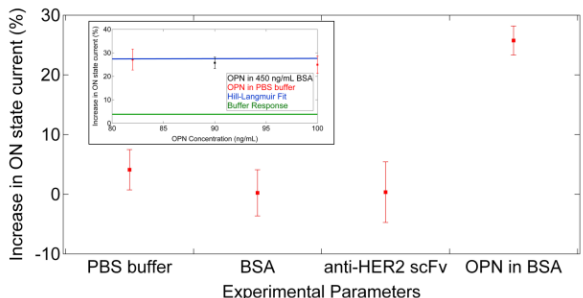

Figure 5: Summary of data from control experiments. Devices exposed to neat PBS buffer showed a response of +4%. Exposure to bovine serum albumin (BSA) at 450 ng/mL gave a null response. Devices prepared with the anti-HER2 scFv antibody in place of anti-OPN scFv and exposed to 90 ng/mL OPN also gave a null response. (Inset) Devices prepared with anti-OPN scFv antibodies and exposed to a mixture of 90 ng/mL OPN and 450 ng/mL BSA background protein gave a response identical to that expected for 90 ng/mL OPN in plain buffer.

The specificity of the sensor was further investigated by exposing 23C3 scFv-functionalized devices to a mixture of OPN (90 ng/mL) and BSA (450 ng/mL) to partially simulate the complexity characteristic of clinical samples. The sensor response of ~ 26% was statistically identical to that observed for devices exposed to pure OPN (see Fig. 5).

## Conclusions

We demonstrated that biosensor devices comprised of high affinity, engineered antibodies coupled to sensitive carbon nanotube transduction elements were capable of detecting prostate cancer biomarker at concentrations as low as 1 pg/mL (30 fM), three orders of magnitude lower than ELISA immunoassays, the current clinical standard[37]. The experiments showed an antigen-specific, concentration-dependent sensor response over a wide range of concentrations (1 pg/ml – 1 µg/ml) that was in excellent quantitative agreement with a model based on the Hill-Langmuire equation of equilibrium thermodynamics. Control experiments indicated that the anti-OPN scFv retains its highly specific binding characteristics when integrated into the hybrid nanostructure: Exposing the sensor to PBS buffer that contained a high concentration of bovine serum albumin did not induce a sensing response, and devices functionalized with 4D5 scFv, which binds specifically to HER2 and not to OPN, did not produce a sensing response upon exposure to OPN. We explored the response of the device to protein mixtures and found that the response to OPN in a concentrated protein background was equal to that measured for OPN in pure buffer. These observations make us optimistic that this device concept may be generalized to many other protein species and is perhaps suitable for translation into a useful tool to help diagnose disease and guide its treatment.

## Methods

**Device Fabrication:** Silicon (p++ doped) wafers with 500 nm thermally grown oxide were covered with randomly dispersed iron nitrate catalyst. Carbon nanotubes were grown at 900°C via chemical vapor deposition with methane feedstock in an argon/hydrogen reducing atmosphere as published previously[15]. Source and drain electrodes separated by 2.5 µm were patterned photolithographically directly on the carbon nanotube network with a bilayer resist process of PMGI and Shipley 1813 followed by metal deposition (5 nm Ti/40nm Pd) in a thermal evaporator. Liftoff in acetone and Microposit CD-26 was followed by a brief anneal in Ar/$H_2$ at 350°C to remove photoresist residues[27]. Device current-gate voltage (I-$V_G$) characteristics were measured under ambient laboratory conditions. Only carbon nanotube transistors with ON/OFF ratios greater than 100 and differences between turn-on voltage and turn-off voltage less than 6 V were used in subsequent measurements.

**Expression and Purification of anti-OPN 23C3 scFv**: A gene encoding the 23C3 scFv, in a Vh-linker-Vl orientation, was synthesized based on the publically available amino acid sequence (PDB ID: 3CXD) and with a codon usage bias for expression in *E. coli*. The gene was cloned into the pSYN2 expression vector as an NcoI/XhoI fragment and the protein was purified to a level of 0.2 mg/L of culture from the periplasmic space of TG1 *E. coli* by sequential Ni-NTA and size exclusion chromatography as previously described[38]. For use as a control, the hu4D5 anti-HER2 scFv corresponding to the FDA-approved mAb trastauzumab, was synthesized from the publically available amino acid sequence in a Vl-linker-Vh orientation as previously described[39], expressed in TG1 *E. coli* and purified as described for 23C3 scFv.

**Functionalization** NTFETs were first functionalized using 4-carboxybenzene diazonium tetrafluoroborate that we synthesized according to a published recipe.[40] Devices were immersed in diazonium salt solution (2.5 mg/mL deionized water) for 1 hour at 40°C to create multiple $sp^3$ hybridized sites along the nanotube ending in a carboxylic acid group[21]. After incubation in a water bath, devices were rinsed in acetone, methanol and DI water. The carboxylic acid groups were then activated and stabilized in a solution of EDC and NHS at concentrations of 6 mg and 16 mg per 15 mL MES buffer respectively for 15 minutes at room temperature followed by a DI water rinse. A solution of scFv antibodies (1 µg/mL) and pH 7.3 was then pipetted onto the devices in a humid environment to keep the solution from evaporating, causing NHS-stabilized sites to be displaced by scFv protein over an incubation period of one hour.

# References

1.     Group, U. S. C. S. W., United States Cancer Statistics: 1999–2007 Incidence and Mortality Web-based Report. U.S. Department of Health and Human Services, C. f. D. C. a. P. a. N. C. I., Ed. Atlanta, 2010.
2.     Thompson, I. M.; Pauler, D. K.; Goodman, P. J.; Tangen, C. M.; Lucia, M. S.; Parnes, H. L.; Minasian, L. M.; Ford, L. G.; Lippman, S. M.; Crawford, E. D.; Crowley, J. J.; Coltman, C. A., Prevalence of prostate cancer among men with a prostate-specific antigen level <= 4.0 ng per milliliter. *New Engl J Med* **2004,** *350* (22), 2239-2246.
3.     Smith, D. S.; Humphrey, P. A.; Catalona, W. J., The early detection of prostate carcinoma with prostate specific antigen - The Washington University experience. *Cancer* **1997,** *80* (9), 1852-1856.
4.     Fedarko, N. S.; Jain, A.; Karadag, A.; Van Eman, M. R.; Fisher, L. W., Elevated serum bone sialoprotein and osteopontin in colon, breast, prostate, and lung cancer. *Clin Cancer Res* **2001,** *7* (12), 4060-6.
5.     Sodek, J.; Ganss, B.; McKee, M. D., Osteopontin. *Crit Rev Oral Biol Med* **2000,** *11* (3), 279-303.

6. Morimoto, J.; Kon, S.; Matsui, Y.; Uede, T., Osteopontin as a target molecule for the teratment of inflammatory diseases. *Curr. Drug Targets* **2010,** *11*, 494 - 505.
7. Bellahcene, A.; Castronovo, V.; Ogbureke, K. U.; Fisher, L. W.; Fedarko, N. S., Small integrin-binding ligand N-linked glycoproteins (SIBLINGs): multifunctional proteins in cancer. *Nat. Rev. Cancer* **2008,** *8*, 212 - 226.
8. Hotte, S. J.; Winquist, E. W.; Stitt, L.; Wilson, S. M.; Chambers, A. F., Plasma osteopontin: associations with survival and metastasis to bone in men with hormone-refractory prostate carcinoma. *Cancer* **2002,** *95* (3), 506-12.
9. Fan, K.; Dai, J.; Wang, H.; Wei, H.; Cao, Z.; Hou, S.; Qian, W.; Li, B.; Zhao, J.; Xu, H.; Yang, C.; Guo, Y., Treatment of collagen-induced arthritis with an anti-osteopontin monoclonal antibody through promotion of apoptosis of both murine and human activated T cells. *Arthritis Rheum* **2008,** *58* (7), 2041-52.
10. Anborgh, P. H.; Wilson, S. M.; Tuck, A. B.; Winquist, E.; Schmidt, N.; Hart, R.; Kon, S.; Maeda, M.; Uede, T.; Stitt, L. W.; Chambers, A. F., New dual monoclonal ELISA for measuring plasma osteopontin as a biomarker associated with survival in prostate cancer: clinical validation and comparison of multiple ELISAs. *Clin Chem* **2009,** *55* (5), 895-903.
11. Plumer, A.; Duan, H.; Subramaniam, S.; Lucas, F. L.; Miesfeldt, S.; Ng, A. K.; Liaw, L., Development of fragment-specific osteopontin antibodies and ELISA for quantification in human metastatic breast cancer. *BMC Cancer* **2008,** *8*, 38.
12. Ward, A. M.; Catto, J. W.; Hamdy, F. C., Prostate specific antigen: biology, biochemistry and available commercial assays. *Ann Clin Biochem* **2001,** *38* (Pt 6), 633-51.
13. Gao, X.; Zheng, G.; Lieber, C. M., Subthreshold regime has the optimal sensitivity for nanowire FET sensors. *Nano Lett.* **2010,** *10*, 547 - 552.
14. Chen, R. J.; Choi, H. C.; Bangsaruntip, S.; Yenilmez, E.; Tang, X. W.; Wang, Q.; Chang, Y. L.; Dai, H. J., An investigation of the mechanisms of electronic sensing of protein adsorption on carbon nanotube devices. *J. Am. Chem. Soc.* **2004,** *126* (5), 1563-1568.
15. Staii, C.; Chen, M.; Gelperin, A.; Johnson, A. T., DNA-decorated carbon nanotubes for chemical sensing. *Nano Lett.* **2005,** *5* (9), 1774-1778.
16. Allen, B. L.; Kichambare, P. D.; Star, A., Carbon nanotube field-effect-transistor-based biosensors. *Adv Mater* **2007,** *19* (11), 1439-1451.
17. Heller, I.; Janssens, A. M.; Mannik, J.; Minot, E. D.; Lemay, S. G.; Dekker, C., Identifying the mechanism of biosensing with carbon nanotube transistors. *Nano Lett* **2008,** *8* (2), 591-595.
18. Star, A.; Gabriel, J. C. P.; Bradley, K.; Gruner, G., Electronic detection of specific protein binding using nanotube FET devices. *Nano Lett.* **2003,** *3* (4), 459-463.
19. Kuang, Z. F.; Kim, S. N.; Crookes-Goodson, W. J.; Farmer, B. L.; Naik, R. R., Biomimetic Chemosensor: Designing Peptide Recognition Elements for Surface Functionalization of Carbon Nanotube Field Effect Transistors. *ACS Nano* **2010,** *4* (1), 452-458.
20. So, H. M.; Park, D. W.; Jeon, E. K.; Kim, Y. H.; Kim, B. S.; Lee, C. K.; Choi, S. Y.; Kim, S. C.; Chang, H.; Lee, J. O., Detection and titer estimation of Escherichia coli using aptamer-functionalized single-walled carbon-nanotube field-effect transistors. *Small* **2008,** *4* (2), 197-201.
21. Goldsmith, B. R.; Mitala, J. J.; Josue, J.; Castro, A.; Lerner, M. B.; Bayburt, T. H.; Khamis, S. M.; Jones, R. A.; Brand, J. G.; Sligar, S. G.; Luetje, C. W.; Gelperin, A.; Rhodes, P. A.; Discher, B.; Johnson, A. T. C., Biomimetic chemical sensors using nanoelectronic readout of olfactory receptor proteins. *ACS Nano* **2011,** *5*, 5408-5416.
22. Kim, S. N.; Rusling, J. F.; Papadimitrakopoulos, F., Carbon nanotubes for electronic and electrochemical detection of biomolecules. *Adv Mater* **2007,** *19* (20), 3214-3228.
23. Cao, Q.; Rogers, J. A., Ultrathin Films of Single-Walled Carbon Nanotubes for Electronics and Sensors: A Review of Fundamental and Applied Aspects. *Adv Mater* **2009,** *21* (1), 29-53.

24. Sanchez, S.; Fabregas, E.; Pumera, M., Detection of biomarkers with carbon nanotube-based immunosensors. *Methods Mol Biol* **2010,** *625*, 227-37.
25. Fan, K.; Zhang, B.; Yang, H.; Wang, H.; Tan, M.; Hou, S.; Qian, W.; Li, B.; Wang, H.; Dai, J.; Guo, Y., A humanized anti-osteopontin antibody protects from Concanavalin A induced-liver injury in mice. *Eur. J. Pharmacol.* **2011,** *657*, 144-151.
26. Holliger, P.; P.J., H., Engineered antibody fragments and the rise of single domains. *Nat. Biotechnol.* **2005,** *23*, 1126 - 1136.
27. Khamis, S. M.; Jones, R. A.; Johnson, A. T. C., Optimized photolithographic fabrication process for carbon nanotube devices. *AIP Advances* **2011,** *1*, 022106.
28. Strano, M. S.; Dyke, C. A.; Usrey, M. L.; Barone, P. W.; Allen, M. J.; Shan, H. W.; Kittrell, C.; Hauge, R. H.; Tour, J. M.; Smalley, R. E., Electronic structure control of single-walled carbon nanotube functionalization. *Science* **2003,** *301* (5639), 1519-1522.
29. Lu, Y.; Lerner, M. B.; Qi, Z. J.; Mitala, J. J.; Lim, J. H.; Discher, B. M.; Johnson, A. T. C., Graphene-protein bioelectronic devices with wavelength-tunable photoresponse. *Appl. Phys. Lett.* **2012,** *100*, 033110.
30. Goldsmith, B. R.; Mitala, J. J.; Josue, J.; Castro, A.; Lerner, M. B.; Bayburt, T. H.; Khamis, S. M.; Jones, R. A.; Brand, J. G.; Sligar, S. G.; Luetje, C. W.; Gelperin, A.; Rhodes, P. A.; Discher, B. M.; Johnson, A. T. C., Biomimetic Chemical Sensors Using Nanoelectronic Readout of Olfactory Receptor Proteins. *Acs Nano* **2011,** *5* (7), 5408-5416.
31. Du, J.; Hou, S.; Zhong, C.; Lai, Z.; Yang, H.; Dai, J.; Zhang, D.; Wang, H.; Guo, Y.; Ding, J., Molecular basis of recognition of human osteopontin by 23C3, a potential therapeutic antibody for treatment of rheumatoid arthritis. *J Mol Biol* **2008,** *382* (4), 835-42.
32. Squires, T. M.; Messinger, R. J.; Manalis, S. R., Making it stick: convection reactdion and diffusion in surface-based biosensors. *Nat. Biotechnol.* **2008,** *26*, 417-426.
33. Arlett, J. L.; Myers, E. B.; Roukes, M. L., Comparative advantages of mechanical biosensors. *Nat. Nanotechnol.* **2011,** *6*, 203-215.
34. Hill, A., The possible effects of the aggregation of the molecules of hemoglobin on its oxygen dissociation curve. *Journal of Physiology* **1910,** *40*, 4-7.
35. Slamon, D. J.; Leyland-Jones, B.; Shak, S.; Fuchs, H.; Paton, V.; Bajamonde, A.; Fleming, T.; Eiermann, W.; Wolter, J.; Pegram, M.; Baselga, J.; Norton, L., Use of chemotherapy plus a monoclonal antibody against HER2 for metastatic breast cancer that overexpresses HER2. *N Engl J Med* **2001,** *344* (11), 783-92.
36. Worn, A.; Pluckthun, A., An intrinsically stable antibody scFv fragment can tolerate the loss of both disulfide bonds and fold correctly. *FEBS Lett* **1998,** *427* (3), 357-61.
37. MacBeath, G., Protein microarrays and proteomics. *Nat Genet* **2002,** *32*, 526-532.
38. Robinson, M. K.; Hodge, K. M.; Horak, E.; Sundberg, A. L.; Russeva, M.; Shaller, C. C.; von Mehren, M.; Shchaveleva, I.; Simmons, H. H.; Marks, J. D.; Adams, G. P., Targeting ErbB2 and ErbB3 with a bispecific single-chain Fv enhances targeting selectivity and induces a therapeutic effect in vitro. *Br J Cancer* **2008,** *99* (9), 1415-25.
39. Kubetzko, S.; Balic, E.; Wiaibel, R.; Zangemeister-Wittke, U.; Pluckthun, A., PEGylation and multimerization of the Anti-p185 HER-2 single chain Fv fragment 4D5. *Journal of Biological Chemistry* **2006,** *281* (46), 35186-35201.
40. Saby, C.; Ortiz, B.; Champagne, G. Y.; Belanger, D., Electrochemical modification of glassy carbon electrode using aromatic diazonium salts .1. Blocking effect of 4-nitrophenyl and 4-carboxyphenyl groups. *Langmuir* **1997,** *13* (25), 6805-6813.